\documentclass[conference,usenatbib]{basi}
\usepackage[british]{babel}
\usepackage{rotating}
\usepackage{dcolumn}
\begin{document}
\title[Hot-Star Models]{Hot-Star Models from 100 to 10,000 \AA}
\author[Claus Leitherer]%
       {Claus Leitherer\thanks{email: \texttt{leitherer@stsci.edu}}\\
       Space Telescope Science Institute, 3700 San Martin Dr., Baltimore, MD 21218, USA}

\pubyear{2011}
\volume{00}
\pagerange{\pageref{firstpage}--\pageref{lastpage}}

\date{Received \today}

\maketitle
\label{firstpage}

\begin{abstract}
The spectral libraries of hot, massive stars which are implemented in the population synthesis code Starburst99
are discussed. Hot stars pose particular challenges for generating libraries. They are
rare, they have an intense radiation field and strong stellar winds, and a luminosity bias
towards ultraviolet wavelengths. These properties require the utilization of theoretical libraries.
Starburst99 uses static non-LTE models at 0.3~\AA\ resolution in the optical, spherically extended,
expanding models at 0.4~\AA\ resolution, in the satellite-ultraviolet, and blanketed, low-resolution
radiation-hydrodynamical models in the extreme ultraviolet down to X-rays. I review the main 
features of each library, compare them to observations, and discuss their link with stellar
evolution models.
\end{abstract}


\section{Introduction}

The past decade has witnessed considerable progress in charting the demographics of high-redshift
galaxies. Panchromatic surveys have established the luminosity functions of star-forming 
galaxies both in the ultraviolet (UV) (Reddy \& Steidel 2009) and in the infrared (IR) and sub-mm (Wardlow et al.
2011). Intermediate dispersion spectroscopy of bright Lyman-break galaxies (LBGs) has been 
pivotal in stellar population trends that cannot be identified from photometric data alone. Such
data can now be routinely obtained for ``normal'' galaxies at redshifts as high as $z \approx 4$, which 
corresponds to an age of the universe of $\sim$900~Myr (Shapley 2011). Owing to the young ages,
the restframe-UV spectra of these galaxies --- redshifted into the optical --- show the signatures of hot, massive stars.
The restframe-UV spectra of LBGs exhibits striking resemblances to those of star-forming galaxies in the local
universe (Schwartz et al. 2006). Both distant and nearby star-forming galaxies and their hot-star
populations are the targets of the spectral libraries discussed in this paper. The stellar contribution to the spectral energy
distribution of these galaxies falls into three categories: (i) the restframe-UV with its distinct stellar-wind
absorption lines from O stars; (ii) the optical/near-IR with numerous weak photospheric absorptions from B and A stars, plus conspicuous
emission lines from H~II regions; (iii) the ionizing extreme UV, which is directly related to the UV wind lines and the
optical nebular lines.
I will use this astrophysical categorization of the spectra of star-forming galaxies to discuss the groups of spectral libraries
of hot stars as implemented in the evolutionary synthesis code Starburst99 (Leitherer et al. 1999; V\'azquez \& Leitherer
2005; Leitherer \& Chen 2009).

\section{Why use Theoretical Libraries?}

Spectral libraries of hot, massive stars must be built from theoretical model atmospheres. This fact is driven not by choice
but by nature. Massive stars are rare and therefore distant --- the closest early-O star $\zeta$~Puppis is at a distance of $\sim$450~pc.
This, together with the concentration of massive stars in the Galactic plane, leads to large dust extinction in the optical/UV
and associated large flux uncertainties. Moreover, large gas column densities produce strong interstellar absorption lines
which can blend stellar features or even fully depress the stellar continuum in the far-UV. Pellerin et al. (2002) used
the Far Ultraviolet Spectroscopic Explorer (FUSE) satellite to collect far-UV spectra of OB stars in the Galaxy, the LMC, and the SMC. Galactic stars even with moderate
dust reddening as low as $E(B-V) = 0.1$ have H$_2$ columns which make large portions of the spectra unusable for library
purposes. Of course, the foremost reason for having to rely on theoretical libraries is the lack of metal-poor, hot, massive stars with
a metal abundance of $Z < \frac{1}{10} Z_\odot$ in
the local universe. Massive stars are short-lived and are forming during an epoch when local star-forming galaxies (including
the Milky Way) have evolved from low to high metallicity. All known metal-poor star-forming galaxies are at distances beyond 
$\sim$10~Mpc, which rules out observations of individual massive stars for the purpose of generating libraries: not only are the 
stars dauntingly faint, but they are located in dense star clusters whose members cannot be resolved even by spectrographs in space. 
This dictates a hybrid approach when building libraries of hot massive stars. We must largely rely on model atmospheres 
but whenever possible, cross-checks and comparisons with observations must be performed to validate this approach. 

Modeling the atmospheres of hot stars is challenging due to severe departures from local thermodynamic equilibrium (LTE)
caused by the intense radiation, low densities, and the presence of supersonic stellar winds initiated by the transfer of 
momentum from the stellar radiation field to the atmospheric plasma. Fortunately, rapid progress during recent years has led 
to an astounding degree of sophistication in the latest generation of models (e.g., Puls 2008). The main challenges are in the 
areas of non-LTE, line-blanketing, and radiatively driven winds. Several groups have independently developed model atmospheres for OB 
stars, with a different emphasis on one or more of these aspects. Plane-parallel models, such as ATLAS (Kurucz 2005) or TLUSTY (Lanz \& Hubeny 2003, 2007), 
cannot account for spectral lines forming in the wind but are suitable for photospheric modeling. The major spherical wind model 
atmospheres are PoW-R (Hamann \& Grä\"afener 2004), PHOENIX (Hauschildt et al. 1999), CMFGEN (Hillier \& Miller 1998), WM-Basic 
(Pauldrach et al. 2001), and FASTWIND (Puls et al. 2005). PoW-R and PHOENIX are optimized for the modeling of Wolf-Rayet (W-R) and late-type stars, 
respectively. FASTWIND's main application is the computation of optical and near-infrared H and He lines. On the other hand, CMFGEN and 
WM-Basic are both widely used for modeling the UV spectra of hot stars. 

Errors due to missing or incorrect atomic data are a concern for the computation of spectra (Hillier 2011). However, this is less
of an issue in hot stars as compared to their cool counterparts where the presence of dust and molecules adds to the complexity
of the modeling. Overall, the state of spectral modeling of {\em hot} stars gives reasonable confidence in the reliability of the
computed spectral libraries.

\section{The Ultraviolet Spectral Region}

The space-UV part of the spectrum (defined as the wavelength region between 912 and 3000~\AA) is dominated by strong stellar-wind lines from
hot massive stars. Massive stars are luminous and develop strong mass loss already when on the main-sequence. The resulting stellar
winds are radiatively driven, with radiative momentum being transferred into kinetic momentum via absorption in
metal lines (Puls, Vink, \& Najarro 2008). Since there is a well-defined relation between the radiative and the kinetic momentum (the
efficiency rate is about 30\%), the kinematic properties of the wind (as observed in the wind lines) reflect the basic stellar 
properties, such as effective temperature $T_{\rm eff}$, luminosity $L$, and gravity $\log g$.

We used WM-Basic for modeling the UV spectra of hot stars (Leitherer et al. 2010). A complete model atmosphere calculation consists of three main blocks: (i) the solution of the hydrodynamics; 
(ii) the solution of the non-LTE-model (calculation of the radiation field and the occupation numbers); (iii) the 
computation of the synthetic spectrum. The three cycles are interdependent and must therefore be solved iteratively.
In the first step, the hydrodynamics is solved for a set of $T_{\rm eff}$, $\log g$, radius $R$, and $Z$, together with an assumed 
radiative line acceleration. In a second step the hydrodynamics is solved by iterating the complete line acceleration 
(which includes the opacities of all important ions) and the temperature and density structure. Afterwards, these structures are again iterated 
together with the line acceleration obtained from the spherical non-LTE model.
The final step consists of the computation of the synthetic spectrum. 

\begin{figure}
\centerline{\includegraphics[width=0.5\textwidth]{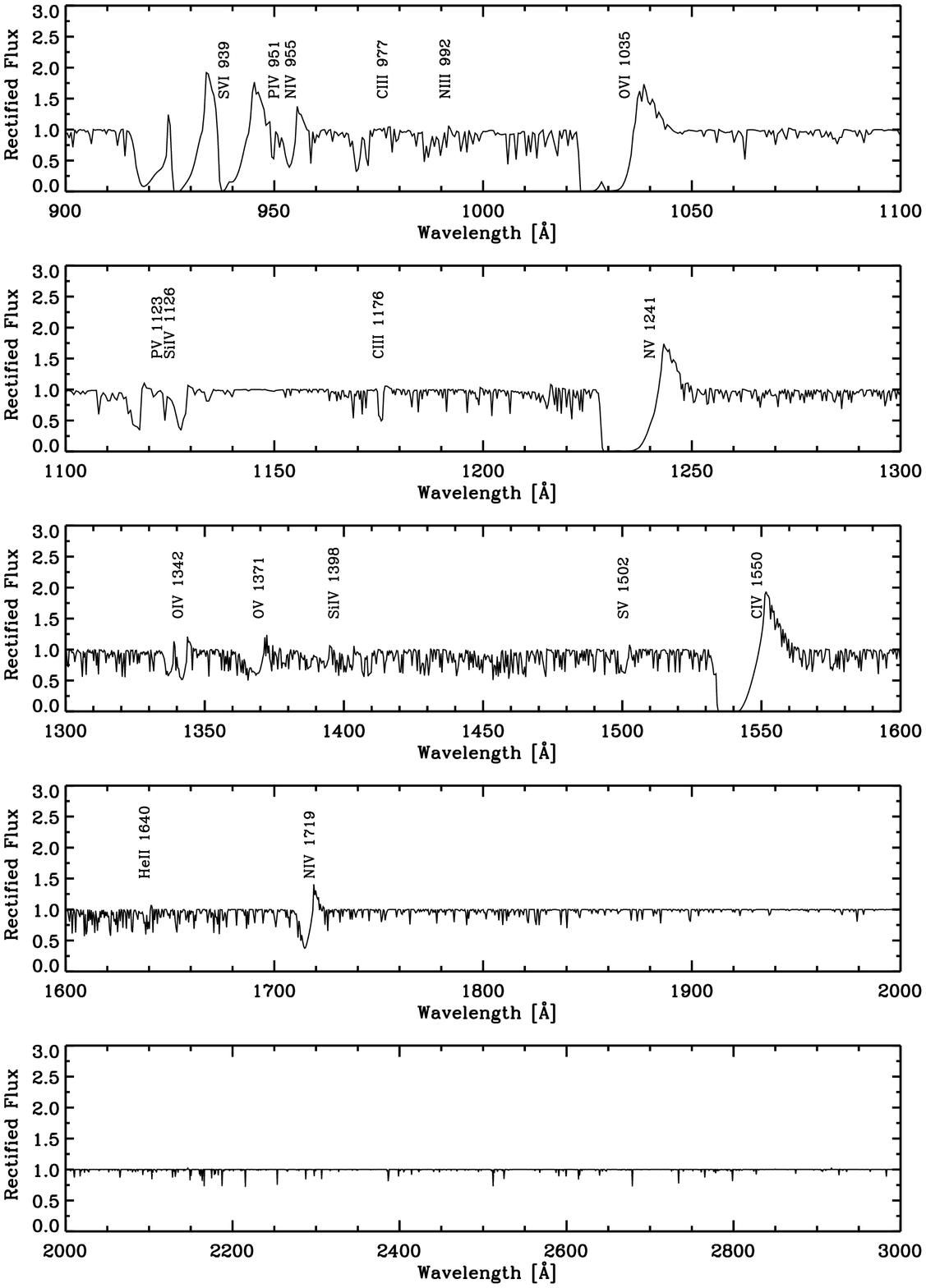}
            \includegraphics[width=0.5\textwidth]{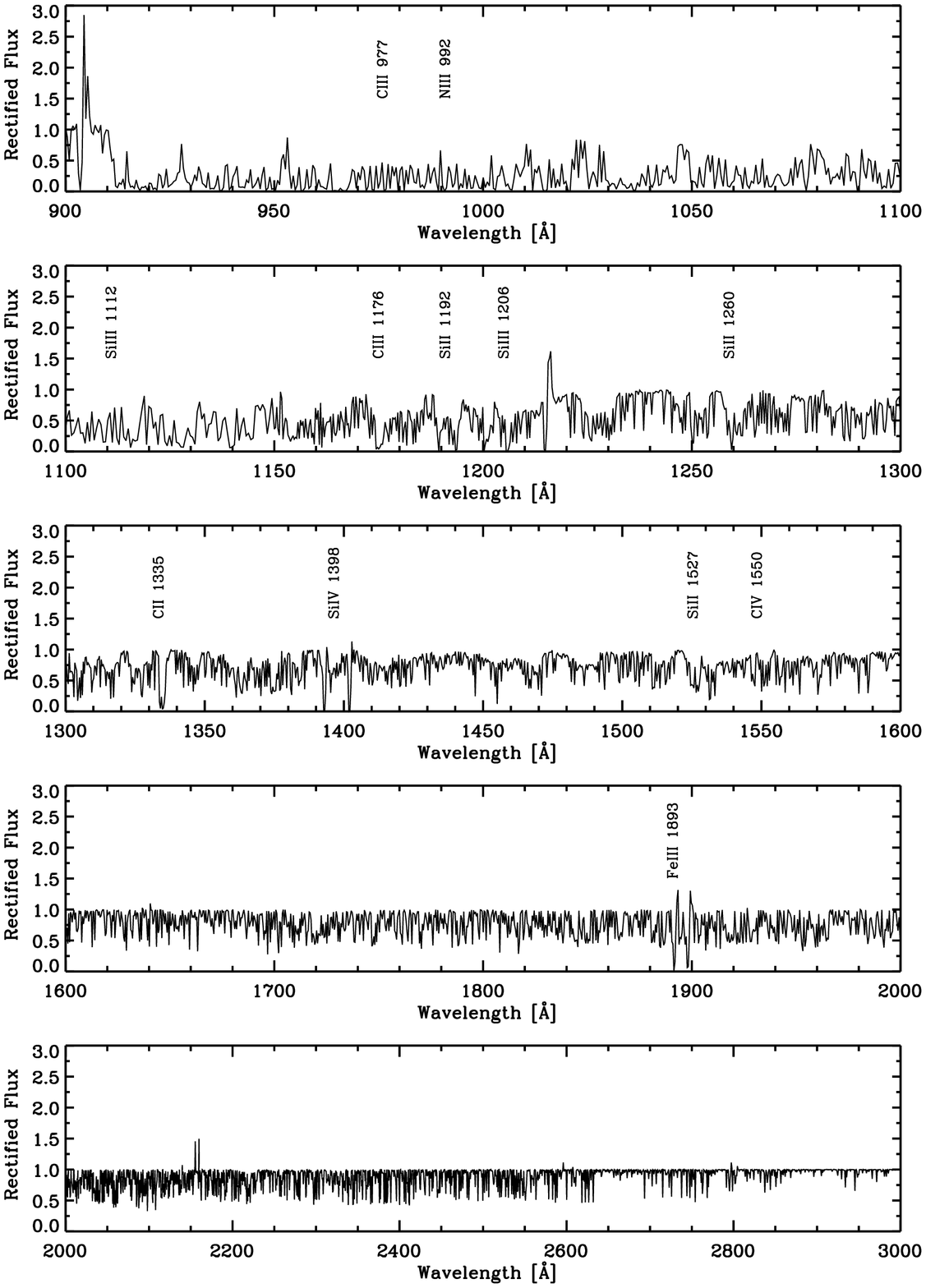}}

\caption{
Examples of theoretical UV spectra. Left: $T_{\rm eff}$ = 37,800 K; $\log g = 3.7$; $Z = Z_\odot$. The corresponding spectral type is O6 III. 
Right: $T_{\rm eff}$ = 15,300 K; $\log g = 1.9$; $Z = Z_\odot$. The corresponding spectral type is B3 Ia. 
The most important spectral lines are identified in both spectra. From Leitherer et al. (2010).\label{hot-cool}}
\end{figure}

Two examples are shown in Fig.~\ref{hot-cool}. The hot star (left panel) has a spectral type of O6~III. Prominent spectral lines are identified. Most of these lines are formed in the stellar wind, as can be seen from the blueshifted absorption components with velocities exceeding $\sim$2000~km~s$^{-1}$. S~VI $\lambda$939, O~VI $\lambda$1035, P~V $\lambda$1123, N~V $\lambda$1240, and C~IV $\lambda$1550 are the strongest features that are uniformly present in O stars. Si~IV $\lambda$1400 is weak in this particular model but can become a strong line in supergiants whose denser winds lead to recombination from Si$^{4+}$ (which is the dominant ionization stage) to Si$^{3+}$ (Walborn \& Panek 1984; Drew 1989). The wavelength region longward of 1800~\AA\ is devoid of both stellar-wind and photospheric lines. This region has little diagnostic value for constraining an O-star population. The O-star spectrum in Fig.~1 can be contrasted with the B-star spectrum in the right panel which corresponds to a B3~Ia supergiant. The high-ionization lines present in the O-star spectrum are much weaker or completely absent in B stars. The strongest lines are, among others, C~III $\lambda$1176, C~II $\lambda$1335, Si~IV $\lambda$1400, and Fe~III $\lambda$1893. This star has a terminal wind velocity of $v_\infty = 300$~km~s$^{-1}$, resulting in comparatively small blueshifts of the absorption components. The line-blanketing by photospheric absorption lines is significant, in particular at the shortest wavelengths where the rectified flux is close to the zero level.

Using this method, we generated a spectral library of 430 spectra for inclusion in evolutionary synthesis models of star-forming galaxies. 
The chosen stellar parameters cover the upper left Hertzsprung-Russell diagram at $L > 10^{2.75} L_\odot$ and $T_{\rm eff} > 20,000$~K. 
The adopted elemental abundances are 0.05~$Z_\odot$, 0.2~$Z_\odot$, 0.4~$Z_\odot$, $Z_\odot$, and 2~$Z_\odot$. The 
spectra cover the wavelength range from 900 to 3000\AA\ and have a resolution of 0.4~\AA. We compared the theoretical spectra to data of individual hot stars in the Galaxy and the Magellanic Clouds obtained with the International Ultraviolet Explorer (IUE) and FUSE satellites and found very good agreement. We then implemented the 
library into the synthesis code Starburst99 where it complements and extends the empirical libraries. 

\begin{figure}
\centerline{\includegraphics[angle=90,width=1.0\textwidth]{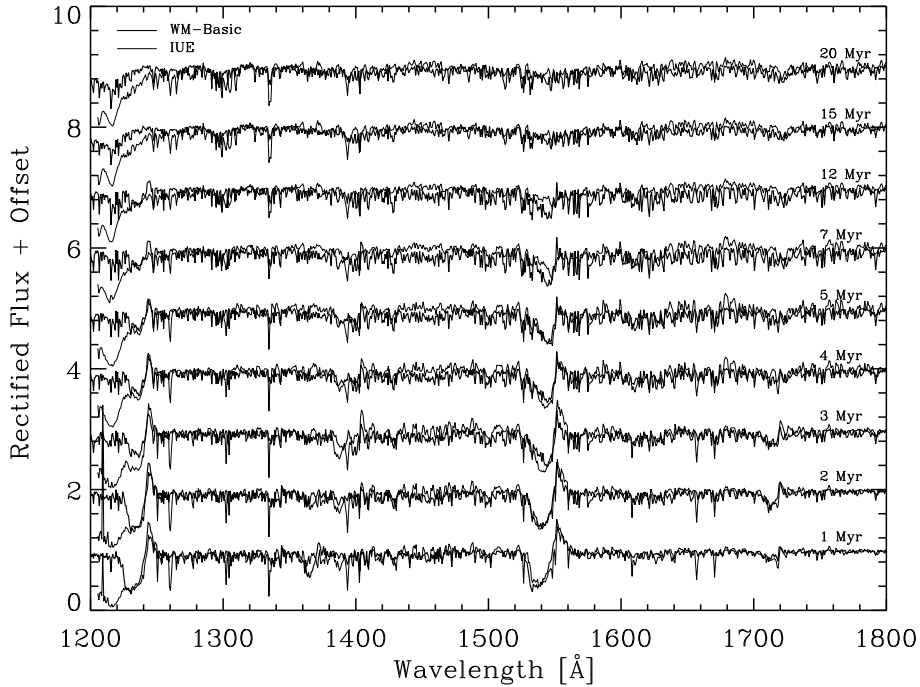}}

\caption{
Comparison of synthetic spectra of a single stellar population obtained with the WM-Basic library with $Z = Z_\odot$ (thick lines) and an 
empirical IUE library of Galactic stars (thin) between 1200 and 1800~\AA. Each spectrum represents one age step from 1 to 20~Myr
as labeled on the right. Salpeter IMF with mass limits 1 and 100~$M_\odot$. From Leitherer et al. (2010).\label{WM-comp}}
\end{figure}

We generated a model series for a standard single stellar population, i.e., an instantaneous burst of star formation following a Salpeter initial mass function (IMF) with mass limits of 1 and 100~$M_\odot$
and following standard evolutionary tracks. In Fig.~\ref{WM-comp}, we show the comparison with the IUE library at $Z_\odot$. The spectral resolution of the WM-Basic library was decreased by a factor of 2 to match that of the IUE library. Plotted are spectra covering ages from 1 to 20~Myr at 9 time steps. 20 Myr is the approximate evolutionary 
time scale of a 10~$M_\odot$ star. Both spectral libraries are complete at (and beyond) this age. Evidently the overall agreement between the spectra generated with the two sets of libraries is excellent for all ages shown in this figure. The age evolution of the population is most clearly reflected in the weakening of the 
N~V $\lambda$1240 and C~IV $\lambda$1550 P~Cygni profiles with time, which results from the decreasing O-star contribution. 
Si~IV $\lambda$1400 exhibits the well-known luminosity effect and is strongest when massive O supergiants appear between 3 and 5 Myr. 
N~IV $\lambda$1720 behaves similarly. Numerous narrow absorption lines in the empirical spectra have interstellar origin. 
Obviously such contamination is absent in the theoretical spectra, thereby greatly facilitating automatic spectral fitting procedures 
that utilize the full spectrum. Notice Si~IV, whose interstellar components contribute significantly to the total profile and 
introduce a major uncertainty in the interpretation of this line. Lyman-$\alpha$ has strong damping wings in the empirical data
due to Galactic H~I absorption. This affects the blue edge of the N~V P~Cygni profile. Again, the library generated with WM-Basic 
does not have this issue. Finally, the normalization of the empirical spectra around 1450~\AA\ is questionable. 
Severe blanketing makes the continuum determination in this wavelength region a challenge and causes the incorrect continuum location in the IUE data.

The new theoretical library is a significant improvement over the existing empirical library and allows the modeling and interpretation of the 
UV spectra of massive star populations for essentially any chemical composition of interest and at any redshift.

\section{The Optical Spectral Region}

The WM-Basic library as implemented into Starburst99 is confined to wavelengths below 3000~\AA. The WM-Basic atmosphere code makes simplifying assumptions which make it not
suitable for calculating the optical line spectra (such as the omission of Stark broadening, which is important for computing the Balmer lines). The
optical spectral region of a star-forming galaxy is weighted towards less massive stars, which allows alternative models to be used. Whereas the UV wind lines are related to O stars with
masses of 20 to 100~$M_\odot$, the optical continuum and line spectrum traces less massive stars of spectral types B and A. Wind effects
are less pronounced in these stars, but non-LTE and line-blanketing are crucial. Consistently treating metal line-blanketing in non-LTE atmospheres 
is challenging but feasible. Line-blanketing has the effect of blocking UV radiation, which then emerges at optical and longer wavelengths.  Some of 
the blocked photons are back-scattered, causing a back-warming effect, which leads to an enhanced ionization in the inner atmosphere. 
This important process has now been incorporated into the most recent generation of models and has led to the realization of a significantly 
revised $T_{\rm eff}$ scale of OB stars (Martins, Schaerer, \& Hillier 2005a).

\begin{figure}
\centerline{\includegraphics[width=0.75\textwidth]{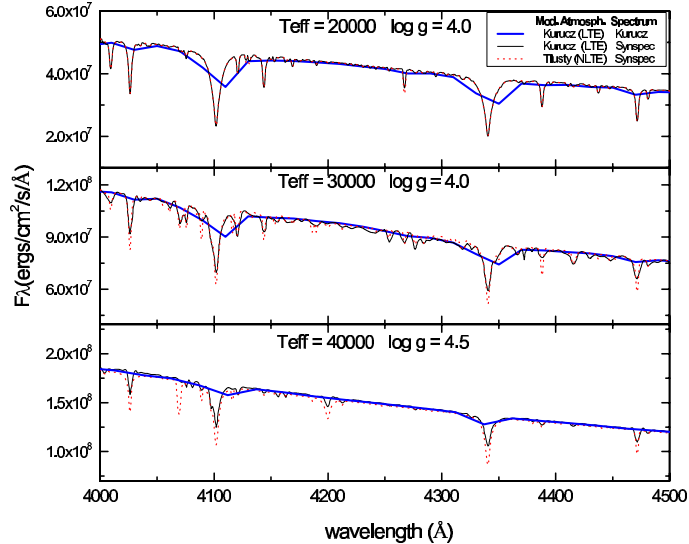}}

\caption{
Comparison between non-LTE and LTE models for solar abundances. Dashed: Kurucz (LTE) at a low resolution of 20~\AA; dotted: Kurucz (LTE) recomputed 
with SYNSPEC at a resolution of 0.3~\AA; solid: TLUSTY (non-LTE) and SYNSPEC at a resolution of 0.3~\AA. The non-LTE models are a dramatic improvement 
over previous model spectra for hotter $T_{\rm eff}$. From Martins et al. (2005b).\label{NLTE}}
\end{figure}

Martins et al. (2005b) computed 1654 high-resolution stellar spectra with a sampling of 0.3~\AA\ and covering the wavelength range from 3000 to 7000~\AA. The grid covers the full Hertzsprung-Russell diagram for chemical abundances of  0.1~$Z_\odot$, 0.5~$Z_\odot$, $Z_\odot$, and 2~$Z_\odot$. The spectra were 
generated using the most appropriate models available for each temperature and gravity value. The library uses non-LTE line-blanketed models 
for hot stars ($T_{\rm eff} > 27,500$~K) and PHOENIX LTE line-blanketed models for cool stars (3000~K~$< T_{\rm eff} <$~4500~K). Kurucz atmosphere models were used
for the parameter space in between. The replacement of the traditional Kurucz atmospheres at the hot and cool end is significant. PHOENIX models account 
for the overall energy distribution in a self-consistent way because of their extensive molecular line list. These molecular transitions are missing in the 
widely used Kurucz models. Lejeune, Cuisinier, \& Buser (1998) provided an empirical correction to the Kurucz models. This correction becomes obsolete with the new fully blanketed models.
Important age diagnostics such as the helium lines are strongly affected by non-LTE effects in OB stars. LTE models predict too weak equivalent widths.  An example of the corresponding 
improvement is shown in Fig.~\ref{NLTE}. The figure suggests that for higher temperatures non-LTE models make a significant difference, as already shown by Hauschildt et al. (1999). LTE models are adequate for B-type stars, whereas for hotter stars non-LTE effects become progressively more important.

Gonz\'alez Delgado et al. (2005) tested this library by linking them to a population synthesis code to calculate high-resolution spectra of a stellar population 
in the 3000~--~7000~\AA\ range. These models predict the indexes of metallic lines and the absorption-line profiles of the hydrogen Balmer series and 
helium lines as a function of age and metallicity for an instantaneous burst. The resulting synthetic models agree very well with observations of 
individual star clusters. Discrepancies are mainly associated with the limitations and uncertainties in stellar evolution models. These limitations can be relevant 
when red supergiants become important in their contribution, in particular at low metallicity.

\section{The Ionizing Ultraviolet Region}

The wavelength range below 912~\AA\ is the source of the radiative heating of the interstellar gas. While this wavelength region is usually not considered part of
a spectral library, it is relevant in that it provides an additional invaluable diagnostic. The stellar radiation field in the extreme UV powers the stellar
winds and is responsible for the strengths and radial velocities of the UV wind lines. Therefore the nebular emission lines in the optical spectrum (which
we use as a proxy for the ionizing radiation field) probe the same energies as the wind lines and should lead to consistent results. 

\begin{figure}
\centerline{\includegraphics[width=1.0\textwidth]{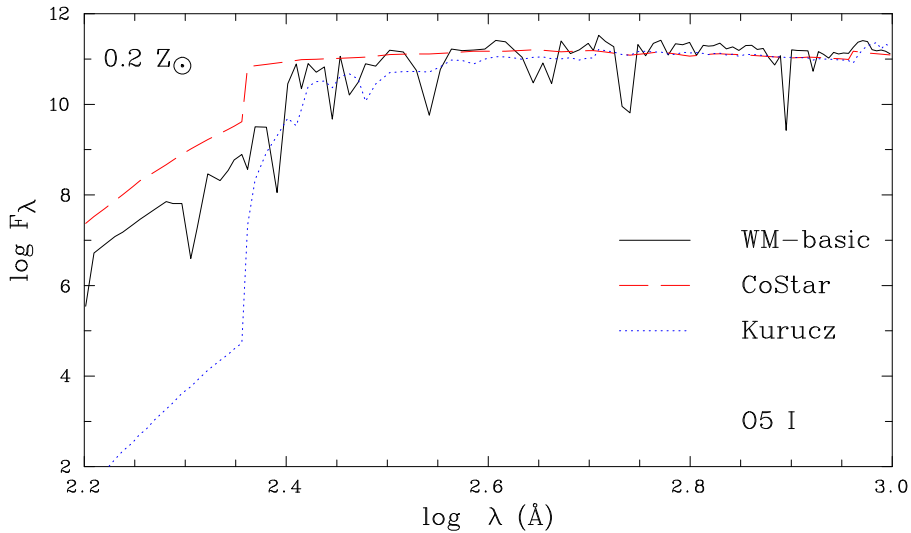}}

\caption{
Comparison of the ionizing UV fluxes for an early-O supergiant from a WM-Basic (solid), a CoStar (dashed), and a Kurucz model (dotted)
at $Z = 0.2 Z_\odot$. From Smith et al. (2002).\label{EUV}}
\end{figure}

The dramatic effects of non-LTE and sphericity effects in the model atmospheres are illustrated in Fig.~\ref{EUV} where WM-Basic 
models are contrasted with classical, static Kurucz LTE atmospheres and with CoStar models (Schaerer \& de Koter 1997), which are similar to WM-Basic models, 
except for the exclusion of line-blanketing. WM-basic is considered to have the most appropriate physical ingredients among the
three models shown and is a viable choice for photoionization modeling. All models agree rather well in the wavelength range 
where the hydrogen ionizing radiation is emitted ($\lambda > 504$~\AA). The number of photons predicted to be emitted in the 
hydrogen Lyman continuum is a robust quantity which has changed by less than 0.3~dex since the classical work of Panagia (1973). 
The behavior of the ionized helium continuum below 228~\AA\ is in sharp contrast to that of hydrogen. Now, wind effects drastically 
alter the spectrum, depending on blanketing and wind density. For the parameters chosen in Fig.~\ref{EUV}, a vast flux excess over Kurucz models is predicted
for WM-Basic models, yet blanketing somewhat lowers the output below that of the CoStar models. The behavior of the neutral helium continuum
at wavelengths between 228 and 504~\AA\ is bracketed by the cases of H and He$^+$. The astrophysical reason for the flux increase of the
extended models is the velocity gradient due to the outflowing matter. The resulting Doppler shift favors less opacity and more free-bound emission.

Dopita et al. (2006) and Levesque, Kewley, \& Larson (2010) used the Mappings III photoionization
models to produce theoretical emission-line spectra of H~II regions and star-forming galaxies with
WM-Basic SEDs as input sources. Their extensive suite of models is intended to facilitate
detailed studies of star-forming galaxies and their interstellar medium properties, but also serves
as an important test of the model atmospheres. The model sets of the two groups of authors are equivalent 
but differ in some important aspects. The Dopita et al. models take the ratio of the 
mass of the central aging star cluster to the gas pressure as the free parameter. Therefore the
ionization parameters becomes a function of age, as appropriate for individual H~II regions. 
To model the spectra of star-forming galaxies as done by Levesque et al., the individual
H~II regions are integrated for each model age, essentially considering star-forming galaxies to
have spectra that consist of contributions from multiple H~II regions at different ages. 

Levesque et al. (2010) examined the agreement between the photoionization models and the spectra
of local star-forming galaxy populations from several large data sets, including the Sloan Digital Sky Survey. In general the models 
do a reasonably good job in reproducing most metallicity-sensitive line ratios. Some lines,
however, require a harder radiation field than predicted by the models.  This suggests
that further systematic changes are required in the stellar population synthesis models to produce the far-UV ionizing
spectra. It is not clear whether the model atmospheres or some other ingredient is to blame.
Currently it appears that the adoption of stellar evolutionary tracks which include the effects
of rotation may provide the necessary hardening of the spectrum (Leitherer \& Ekstr\"om 2012).

\section{Applications}

As a demonstration of the library's capabilities I am giving two recent applications, one
related to the chemical composition of high-redshift galaxies, and the other to the IMF of local star-forming galaxies.

A major challenge in deriving abundances from nebular emission lines is that the
strongest and best calibrated features are all located in the rest-frame optical, which
is redshifted into the near-IR at cosmological redshifts. Observations in this wavelength regime from the ground
are plagued by high (and variable) sky background and by the lack
of multi-object spectrographs. There is therefore a strong incentive to test
abundance measures based on the spectral features from massive stars which dominate
the rest-frame UV spectra of high-z galaxies.

Initial applications of this technique turn out to be quite promising 
(Rix et al. 2004; Halliday et al. 2008).  A spectrum created
by adding together 75 star-forming galaxy spectra at $z \approx 2$ from the
GMASS survey was used to estimate the iron abundance by measuring the strengths of
weak photospheric absorption lines. A Starburst99 model with a stellar chemical
composition of $\sim$20\% solar was found to match the data. This abundance is 
similar to those derived in individual LBGs, such as the proto-type 
MS1512-cB58 (a strongly lensed LBG at
$z = 2.73$), for which a stellar metallicity of 0.4~$Z_\odot$ was found. The
latter value is in excellent agreement with the metallicity of the
interstellar medium of this galaxy. This method can provide an additional way to determine the
metallicity of high-z galaxies, particularly at redshifts $z > 4$ where emission-line diagnostics
are difficult to apply from the ground. 

The UV lines are useful for probing the IMF. The lines depend on the
wind properties, which in turn depend on the stellar luminosity $L$. Since there
is a well established mass-luminosity relation in stars, the wind lines trace stellar mass
and ultimately the IMF of the population (Leitherer 2011). This is
equivalent to the nebular H$\alpha$ method for determining the IMF but circumvents the interstellar radiative
transfer in gas and dust because the wind absorption occurs in the outskirts
of the star itself. The only free parameters
in the modeling are the star-formation history of the population and the IMF.
Since massive stars have evolutionary timescales of less than 0.1~Gyr, star-formation 
equilibrium is reached quickly, and the population properties become
time-independent in statistically large systems. This essentially leaves the IMF
as the only free parameter.

\begin{figure}
\centerline{\includegraphics[angle=90,width=1.0\textwidth]{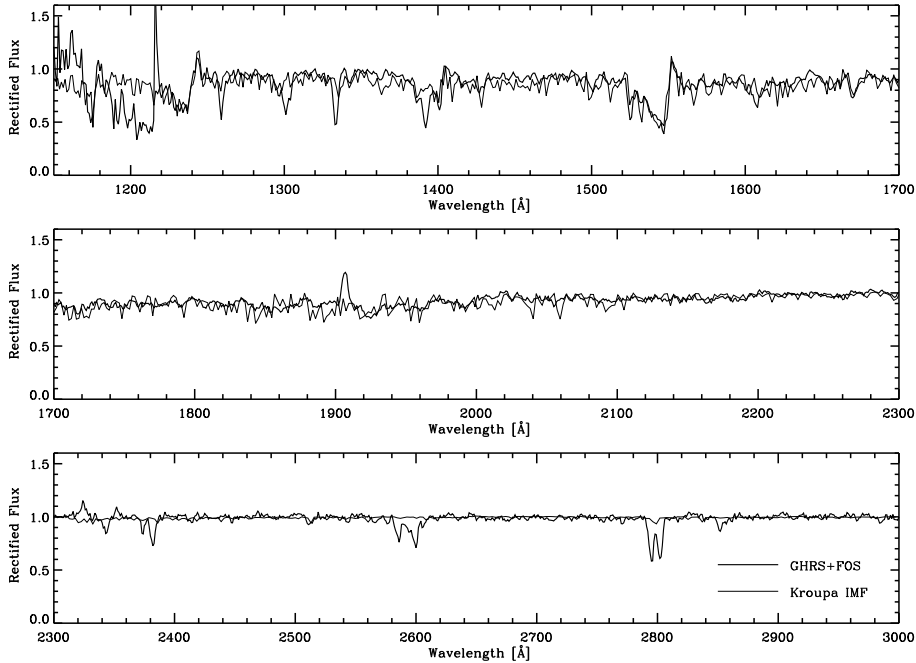}}

\caption{
Comparison between the average spectrum of 46 local star-
forming regions generated by Leitherer et al. (2011; thick line) and a 
synthetic spectrum with a Kroupa/Salpeter IMF populated with stars up to
$M = 100~M_\odot$ (thin line). The narrow absorption lines (e.g., at 1260~\AA,
1335~\AA, 2600~\AA, 2800~\AA) are interstellar and absent in the synthetic spectrum. 
There is strong geocoronal and Milky Way Lyman-$\alpha$ contamination in the
data. Note the excellent agreement for the wind features of N V $\lambda$1240 and C IV $\lambda$1550.\label{IMF}}
\end{figure}

Fig.~\ref{IMF} is a comparison between an average UV spectrum of local star-forming galaxies
and a standard model using the theoretical UV library.  The observational high-S/N spectrum 
is the mean of 46 individual spectra. The average oxygen abundance of log(O/H) + 12 = 8.5 is close
to, but slightly below the solar value. The synthetic UV spectrum was generated
for a standard equilibrium population at solar chemical composition following a
Kroupa IMF between 0.1 and 100~$M_\odot$. (Note that this IMF is identical to the
Salpeter IMF for massive stars.)  The key constraints are the N V, Si IV, and C IV lines.
The interpretation of the Si IV doublet requires care because of the presence of
additional strong interstellar lines which are obviously absent in the model. IMF slopes steeper 
than $\sim$2.6 and flatter than $\sim$2.0 can be excluded. There are no
additional adjustable parameters.

\section{The Road Ahead}

Theoretical model atmospheres of hot stars have reached a stage of maturity to allow
the generation of astrophysically useful stellar library spectra. While comparison with data gives
confidence in the models, I would like to end this contribution by mentioning known deficiencies and
hurdles. 

A major issue plaguing the computed UV lines concerns wind inhomogeneities. The models
discussed here assume smooth winds, yet the ouflows are known to have significant density
fluctuations (Puls et al. 2008). This causes some lines, in particular those from ions far from
the dominant ionization stage, to be discrepant. An example is the O~V $\lambda$1371 line, which
is predicted to be present in very hot O stars but in reality is hardly ever observed. The reason
is understood: on average, a larger mass fraction of the winds is confined to higher density than
assumed in the models (which are clump-free). This shifts the average ionization balance to lower
stages and drastically weakens O~V (Bouret et al. 2003). Efforts to account for clumped winds 
are underway (Sundqvist et al. 2011) and will hopefully allow us to address one of the most
important deficiencies of the current spectral library.

Completeness is an issue for almost every spectral library. In the case of hot stars,
the challenges are both at the low- and at the high-mass end. At low masses, it is important
to realize that {\em evolved} hot stars are not included. While such stars are not relevant
in the presence of hot massive stars, they may become a significant contributor to the blue/UV 
luminosity once massive stars disappear in an ageing stellar population (Flores-Fajardo et al.
2011). Theoretical efforts are still lagging in accounting for these stars. At high masses, 
we have to deal with Wolf-Rayet stars whose peculiar evolutionary states, powerful stellar
winds, and anomalous chemical abundances pose formidable challenges. Wolf-Rayet stars, while
luminous, are often believed to be too few in number to be significant for inclusion in
spectral libraries. However, Crowther et al. (2010) demonstrated that the Wolf-Rayet-like
stars in the center of the 30~Doradus Nebula in the LMC have masses possibly reaching up to
300~$M_\odot$. Therefore these stars make a disproportionate contribution to the global 
UV-light budget and must be accounted for in spectral libraries.

\section*{Acknowledgements}

I wish to thank the organizers for local financial support and the STScI DDRF/JDF for travel funds.


\begin{thebibliography}{}

\bibitem[Bouret et al.(2003)]{2003ApJ...595.1182B} Bouret, J.-C., Lanz, T., 
Hillier, D.~J., et al.\ 2003, ApJ, 595, 1182 

\bibitem[Crowther et al.(2010)]{2010MNRAS.408..731C} Crowther, P.~A., 
Schnurr, O., Hirschi, R., et al.\ 2010, MNRAS, 408, 731 

\bibitem[Drew(1989)]{1989ApJS...71..267D} Drew, J.~E.\ 1989, ApJS, 71, 267 

\bibitem[Dopita et al.(2006)]{2006ApJS..167..177D} Dopita, M.~A., Fischera, 
J., Sutherland, R.~S., et al.\ 2006, ApJS, 167, 177 

\bibitem[Flores-Fajardo et al.(2011)]{2011MNRAS.415.2182F} Flores-Fajardo, 
N., Morisset, C., Stasi{\'n}ska, G., 
\& Binette, L.\ 2011, MNRAS, 415, 2182 

\bibitem[Gonz{\'a}lez Delgado et al.(2005)]{2005MNRAS.357..945G} 
Gonz{\'a}lez Delgado, R.~M., Cervi{\~n}o, M., Martins, L.~P., Leitherer, 
C., \& Hauschildt, P.~H.\ 2005, MNRAS, 357, 945 

\bibitem[Halliday et 
al.(2008)]{2008A&A...479..417H} Halliday, C., Daddi, E., Cimatti, A., et al.\ 2008, A\&A, 479, 417 

\bibitem[Hamann 
\& Gr{\"a}fener(2004)]{2004A&A...427..697H} Hamann, W.-R., \& Gr{\"a}fener, G.\ 2004, A\&A, 427, 697 

\bibitem[Hauschildt et al.(1999)]{1999ApJ...525..871H} Hauschildt, P.~H., 
Allard, F., Ferguson, J., Baron, E., 
\& Alexander, D.~R.\ 1999, ApJ, 525, 871 

\bibitem[Hillier(2011)]{2011Ap&SS.336...87H} Hillier, D.~J.\ 2011, Ap\&SS, 336, 87 

\bibitem[Hillier 
\& Miller(1999)]{1999ApJ...519..354H} Hillier, D.~J., \& Miller, D.~L.\ 1999, ApJ, 519, 354 

\bibitem[Kurucz(2005)]{2005MSAIS...8...14K} Kurucz, R.~L.\ 2005, Mem. Soc. Astron. Ital. Suppl., 8, 14 

\bibitem[Lanz 
\& Hubeny(2003)]{2003ApJS..146..417L} Lanz, T., \& Hubeny, I.\ 2003, ApJS, 146, 417 

\bibitem[Lanz 
\& Hubeny(2007)]{2007ApJS..169...83L} ------. 2007, ApJS, 169, 83 

\bibitem[Leitherer(2011)]{2011ASPC..440..309L} Leitherer, C.\ 2011, in UP2010: 
Have Observations Revealed a Variable Upper End of the Initial Mass 
Function, eds. M. Treyer et al. (San Francisco: ASP), 309 

\bibitem[Leitherer 
\& Chen(2009)]{2009NewA...14..356L} Leitherer, C., \& Chen, J.\ 2009, New Astronomy, 14, 356 

\bibitem[Leitherer 
\& Ekstrom(2011)]{2011arXiv1111.5204L} Leitherer, C., \& Ekstrom, S.\ 2012, in IAU Symp. 284, The Spectral 
     Energy Distribution of Galaxies, eds. R. J. Tuffs \& C. C. Popescu (Cambridge: CUP), in press, arXiv:1111.5204 

\bibitem[Leitherer et al.(2010)]{2010ApJS..189..309L} Leitherer, C., Ortiz 
Ot{\'a}lvaro, P.~A., Bresolin, F., et al.\ 2010, ApJS, 189, 309 

\bibitem[Leitherer et al.(1999)]{1999ApJS..123....3L} Leitherer, C., 
Schaerer, D., Goldader, J.~D., et al.\ 1999, ApJS, 123, 3 

\bibitem[Leitherer et al.(2011)]{2011AJ....141...37L} Leitherer, C., 
Tremonti, C.~A., Heckman, T.~M., \& Calzetti, D.\ 2011, AJ, 141, 37 

\bibitem[Lejeune et 
al.(1998)]{1998A&AS..130...65L} Lejeune, T., Cuisinier, F., \& Buser, R.\ 1998, A\&AS, 130, 65 

\bibitem[Levesque et al.(2010)]{2010AJ....139..712L} Levesque, E.~M., 
Kewley, L.~J., \& Larson, K.~L.\ 2010, AJ, 139, 712 

\bibitem[Martins et 
al.(2005)]{2005A&A...436.1049M} Martins, F., Schaerer, D., \& Hillier, D.~J.\ 2005a, A\&A, 436, 1049 

\bibitem[Martins et al.(2005)]{2005MNRAS.358...49M} Martins, L.~P., 
Gonz{\'a}lez Delgado, R.~M., Leitherer, C., Cervi{\~n}o, M., 
\& Hauschildt, P.\ 2005b, MNRAS, 358, 49 

\bibitem[Panagia(1973)]{1973AJ.....78..929P} Panagia, N.\ 1973, AJ, 78, 
929 

\bibitem[Pauldrach et 
al.(2001)]{2001A&A...375..161P} Pauldrach, A.~W.~A., Hoffmann, T.~L., \& Lennon, M.\ 2001, A\&A, 375, 161 


\bibitem[Pellerin et al.(2002)]{2002ApJS..143..159P} Pellerin, A., 
Fullerton, A.~W., Robert, C., et al.\ 2002, ApJS, 143, 159 

\bibitem[Puls(2008)]{2008IAUS..250...25P} Puls, J.\ 2008, in Massive Stars as Cosmic Engines (IAU Symp. 250), 
ed. F. Bresolin, P. A. Crowther, \& J. Puls (Cambridge: CUP), 25 

\bibitem[Puls et 
al.(2005)]{2005A&A...435..669P} Puls, J., Urbaneja, M.~A., Venero, R., et al.\ 2005, A\&A, 435, 669 

\bibitem[Puls et 
al.(2008)]{2008A&ARv..16..209P} Puls, J., Vink, J.~S., \& Najarro, F.\ 2008, A\&ARv, 16, 209 

\bibitem[Reddy 
\& Steidel(2009)]{2009ApJ...692..778R} Reddy, N.~A., \& Steidel, C.~C.\ 2009, ApJ, 692, 778

\bibitem[Rix et al.(2004)]{2004ApJ...615...98R} Rix, S.~A., Pettini, M., 
Leitherer, C., et al.\ 2004, ApJ, 615, 98  

\bibitem[Schaerer 
\& de Koter(1997)]{1997A&A...322..598S} Schaerer, D., \& de Koter, A.\ 1997, A\&A, 322, 598 

\bibitem[Schwartz et al.(2006)]{2006ApJ...646..858S} Schwartz, C.~M., 
Martin, C.~L., Chandar, R., et al.\ 2006, ApJ, 646, 858 

\bibitem[Shapley(2011)]{2011ARA&A..49..525S} Shapley, A.~E.\ 2011, ARA\&A, 49, 525 

\bibitem[Smith et al.(2002)]{2002MNRAS.337.1309S} Smith, L.~J., Norris, 
R.~P.~F., \& Crowther, P.~A.\ 2002, MNRAS, 337, 1309 

\bibitem[Sundqvist et 
al.(2011)]{2011A&A...528A..64S} Sundqvist, J.~O., Puls, J., Feldmeier, A., \& Owocki, S.~P.\ 2011, A\&A, 528, A64 

\bibitem[V{\'a}zquez 
\& Leitherer(2005)]{2005ApJ...621..695V} V{\'a}zquez, G.~A., \& Leitherer, C.\ 2005, ApJ, 621, 695 

\bibitem[Walborn 
\& Panek(1984)]{1984ApJ...280L..27W} Walborn, N.~R., \& Panek, R.~J.\ 1984, ApJ, 280, L27 


\bibitem[Wardlow et al.(2011)]{2011MNRAS.415.1479W} Wardlow, J.~L., Smail, 
I., Coppin, K.~E.~K., et al.\ 2011, MNRAS, 415, 1479 


\end{thebibliography}
\end{document}